\documentclass[useAMS,usenatbib,usegraphicx]{mn2e}

\newcommand{\beq}{\begin{equation}}
\newcommand{\beqa}{\begin{eqnarray}}
\newcommand{\eeq}{\end{equation}}
\newcommand{\eeqa}{\end{eqnarray}}

\title[A Cosmic Relation between Extinction and Star Formation]
{A Cosmic Relation between Extinction and Star Formation}

\author[Oren Zoran, Rennan Barkana and Rodger I. Thompson]{Oren Zoran$^{1}$,
Rennan Barkana$^{1}$ and Rodger I. Thompson$^{2}$
\thanks{E-mail:
Oren\_Zoran@amat.com (OZ); barkana@wise.tau.ac.il (RB);
thompson@as.arizona.edu (RT)}\\ $^{1}$School of Physics and Astronomy,
The Raymond and Beverly Sackler Faculty of Exact Sciences,\\ Tel Aviv
University, Tel Aviv 69978, ISRAEL\\ $^{2}$ Steward Observatory,
University of Arizona, 933 North Cherry Avenue, Tucson, AZ 85721, USA}

\begin{document}

\pagerange{\pageref{firstpage}--\pageref{lastpage}} \pubyear{2005}

\maketitle

\label{firstpage}

\begin{abstract}
We study the relation between the star formation intensity of galaxies
and the extinction by dust of their emitted light. We employ a
detailed statistical analysis of Hubble Deep Field North data to show
a clear positive correlation between the extinction and star formation
intensity at all epochs from redshift 0.4 to 6.5. The extinction
evidently increases with time for a given star formation intensity,
consistent with the expected increase in the metallicity with
time. Our observational results are well fitted at all epochs by a
double power-law model with a fixed shape that simply shifts with
redshift. The correlation between the extinction and the star
formation intensity can be interpreted by combining two other trends:
the correlation between the star formation rate and the gas content of
galaxies, and the evolution of the dust-to-gas ratio in galaxies. If
we assume that Kennicutt's observed relation for the former is valid
at each redshift, then our findings imply an interesting variation in
the dust-to-gas ratio in galaxies within each epoch and with time, and
suggest new ways to investigate the cosmic evolution of this quantity.
\end{abstract}


\section{Introduction}

In this Letter we investigate the relation between the extinction and
the star formation rate (SFR) in galaxies at various
epochs. Understanding this relation is a key for better understanding
the physical conditions in the interstellar medium (ISM) of galaxies
as well as the environments of galaxies and their cosmic
evolution. The SFR in galaxies has been studied extensively in recent
years. The most widely used star formation law, originally conjectured
by \citet{s59}, correlates the SFR in galaxies and their gas density
as a power law. As summarized and integrated by
\citet{k98}, many empirical studies have tested and proven the 
effectiveness of the Schmidt law. Now, extinction by dust is related
to star formation since dust forms from heavy elements ejected during
stellar mass loss. Also, the dust-to-gas ratio is related to the
metallicity of the interstellar gas \citep{sb93,km96}.

It is thus clear that a correlation should exist between the SFR and
the extinction. Indeed, previous observations indicate such a
correlation within galaxies, and for galaxies at various redshifts
[e.g., \citet{as00, cal01, c05}]. The goal of this paper is to observe
this correlation and determine its form for a broad range of
redshifts, and to demonstrate that our conclusions are not strongly
affected by selection effects. In the following section we describe
the dataset used for the analyses in this Letter. In section~3 we test
the dataset to verify that it is statistically valid and can be used
for the analyses to follow. We present our main results and model fits
in sections~4 and 5, and give our conclusions in section~6. We note
that we focus on the correlation of extinction with the star formation
intensity (SFI) of galaxies (which is the SFR per unit area), but
consider in \S~5 also the correlation with the total SFR.

\section{The Data}

We used the dataset described in detail in Thompson, Waymann \&
Storrie-Lombardi (2001; hereafter TWS) and \citet{t03}. It is a
combination of observations of the Hubble Deep Field North (HDF-N)
with the Wide Field Planetary Camera 2 [WFPC 2; \citet{w96}] and with
the Near-Infrared Camera and Multi-Object Spectrometer [NICMOS;
\citet{d00}). The main output of the photometric analysis in TWS
and \citet{t03} was the most likely combination of redshift,
extinction and spectral energy distribution (SED) for each of the
galaxies that passed the selection criteria. However, in this present
work we utilized not only the most likely combinations but the entire
database of the photometric analysis. The idea is that while the
uncertainties are large from each photometric fit to an individual
galaxy, if we include these uncertainties properly then the total
dataset of thousands of galaxies will have strong statistical power
for studying galaxy evolution.

The dataset includes two three-dimensional matrices for each of the
1972 galaxies that passed the selection criteria and were used for the
analysis. The matrix dimensions are $101 \times 51 \times 15$,
corresponding to 101 possible redshift values between 0 and 8, 15
extinction values ranging from $E(B-V)=0$ to 1.0, and 51 possible SED
templates. The first matrix holds the $\chi^2$ values for each of the
redshift, extinction and SED template combinations, determined as
described in TWS by computing the goodness-of-fit to the six observed
photometric fluxes. Note that in computing the $\chi^2$, TWS added a
term to the background noise that is $10\%$ of the flux, to represent
various systematic effects. As discussed further in TWS, this term may
overestimate the noise somewhat and lead to an underestimate of the
$\chi^2$ values, producing in our analysis below a slightly broader
distribution of probable parameter values. The second matrix in the
dataset holds the UV flux at 1500\AA\ rest-frame that corresponds to
each of the redshift, extinction and template combinations. We used
this matrix to determine the SFR that corresponds to the different
combinations according to the broadband 1500\AA\ relation given by
\citet{mpd98}, modified for a
\citet{s98} IMF:
\beq F_{1500} = 10.0 \times 10^{27}\ {\rm SFR}\left[M_{\odot}
{\rm yr} ^{-1} \right]\ {\rm erg\, s^{-1}\, Hz}^{-1}\
. \label{eq:FUV}\eeq
   
In addition to the above photometric dataset, for 131 galaxies
spectroscopic redshifts are available from \citet{c01}. For these
galaxies, we adopted the spectroscopic redshift and used the above
matrices to determine the likelihood of various extinctions and SED
templates given the known redshift.

\section{Testing the Dataset using Redshift Error Analysis}    

\begin{figure}
\includegraphics[width=84mm]{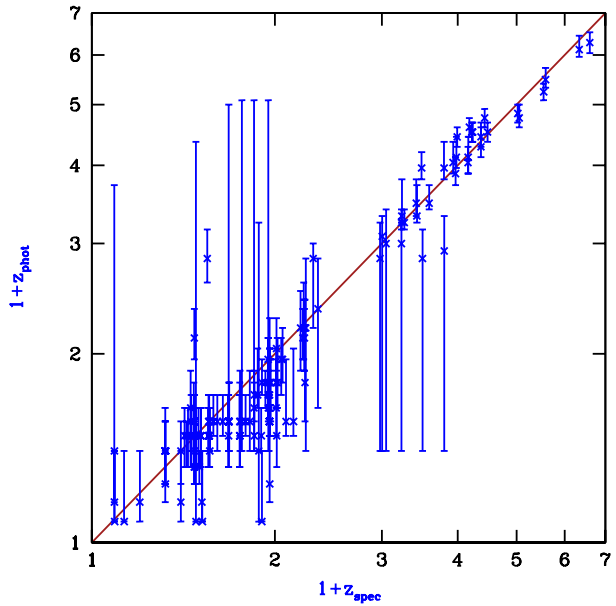}
\caption{Photometric redshift versus spectroscopic redshift for each
galaxy (in a log plot of $1+z$). The confidence limits of 1-$\sigma$
up and 1-$\sigma$ down for the photometric redshifts were determined
using the probability distribution of the photometric redshift of each
galaxy. The spectroscopic data are taken from \citet{c01}}
\label{fig:zcomp}
\end{figure}

In general, a $\chi^2$ value is related to the corresponding
probability through the relation:
\beq P \propto \exp\left(- \frac{1}{2} \chi^2 \right)\ . 
\label{eq:Pi} \eeq
We therefore used this relation to convert, for each galaxy, the
$\chi^2$ values that correspond to the different combinations of
redshift, extinction and template, into their corresponding
probabilities. The combinations for each galaxy were then divided into
two groups, one with redshift values higher than the redshift value of
the most probable combination, and the other with redshift values that
are lower than the redshift value of the most probable combination. We
then found 1-$\sigma$ redshift confidence limits upwards and downwards
by identifying the redshift values at 0.683 of the cumulative
probability of each of the two groups. We emphasize that a full
probability distribution of possible parameters is determined by
fitting to {\it each}\/ observed galaxy. Figure~\ref{fig:zcomp}
compares the photometric redshift values with errors to the
spectroscopic values for the 131 galaxies for which spectroscopic data
are available from \citet{c01}. This plot is similar to the plot in
figure~3 in \citet{t03} with the significant difference that here the
1-$\sigma$ confidence limits take into account the entire distribution
of possible photometric redshifts, a distribution that in some cases
is very wide or multiply-peaked.

\begin{figure}
\includegraphics[width=84mm]{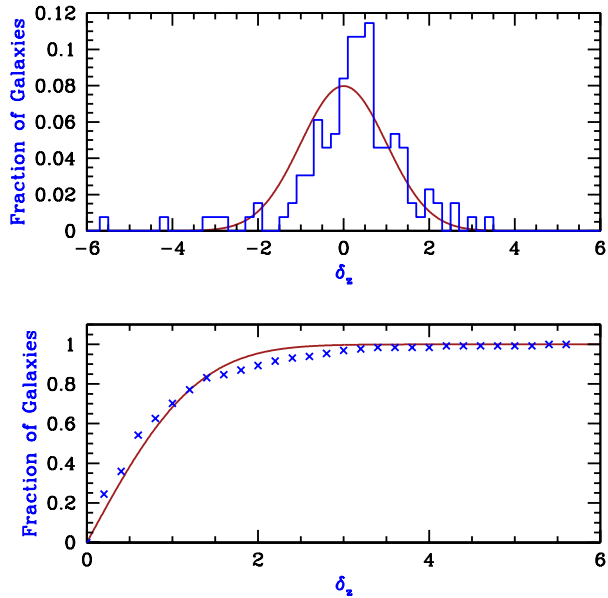}
\caption{Upper panel: The normalized distribution of the
fractional redshift error expressed in standard deviations,
$\delta_z$, defined in equation~(\ref{eq:delz}). Lower panel: The
cumulative distribution of $\delta_z$. The solid curves in the two
panels represent the theoretical Gaussian and error function curves,
respectively.}
\label{fig:zdist}
\end{figure}

The photometric redshifts and their calculated errors, together with
the spectroscopic redshift data, can be used to test the validity of
the photometric analysis dataset. This test is important as the
analysis in the following sections strongly relies on the statistical
properties of the $\chi^2$ matrices. The photometric redshifts deviate
from the spectroscopic ones, but we now consider how these deviations
are distributed. If the statistical analysis is valid, then the
distribution of the deviations between the photometric and the
spectroscopic redshift should be roughly Gaussian, with standard
deviation as predicted by the statistical analysis. Rather than
restricting to a symmetric distribution, we apply the
separately-determined upper and lower confidence
limits. Figure~\ref{fig:zdist} shows the normalized relative redshift
error distribution (upper panel). We define $\delta_z$ as the
fractional redshift error expressed in standard deviations:
\beq \delta_z \equiv \frac{|z_{\rm spec} - z_{\rm phot}|}  
{\sigma_{\pm}}\ , \label{eq:delz} \eeq where $\sigma_{\pm}$ is the
upper or lower confidence limit which is chosen according to whether
the spectroscopic redshift is higher or lower than the photometric
one, respectively. In the Figure we also show (lower panel) the
cumulative distribution of all the points, both those with positive
and those with negative errors. We find that the empirical
distribution is indeed close to a normal distribution with a standard
deviation that agrees well with that predicted by our fitting method;
in particular, the cumulative 1-$\sigma$, 2-$\sigma$ and 3-$\sigma$
values are 0.702, 0.893 and 0.969, compared to the theoretical
expectation of 0.683, 0.954, and 0.997, respectively.
    
\section{Correlating the Extinction with the Star Formation Intensity}

The SFI (denoted $x$) is defined as the SFR per area, expressed in
solar mass per year per proper kiloparsec squared. As described in
section~2, various combinations of redshift, extinction and template
may be associated with each galaxy and for each of these combinations
the SFI can be determined. First the SFR of each combination is
determined from equation~(\ref{eq:FUV}), and then the area is
calculated (for each possible redshift) from the number of pixels that
the galaxy covers in the HDF-N image. The average SFI of the galaxy is
then the SFR divided by the proper area. We emphasize that in the
analysis below we use the average extinction and SFI of each galaxy,
and do not consider individual pixels separately. In this work we
assume in all cosmological conversions a flat cosmology with $h=0.72$,
$\Omega_m=0.27$ and $\Omega_\Lambda=0.73$.

\begin{figure}
\includegraphics[width=84mm]{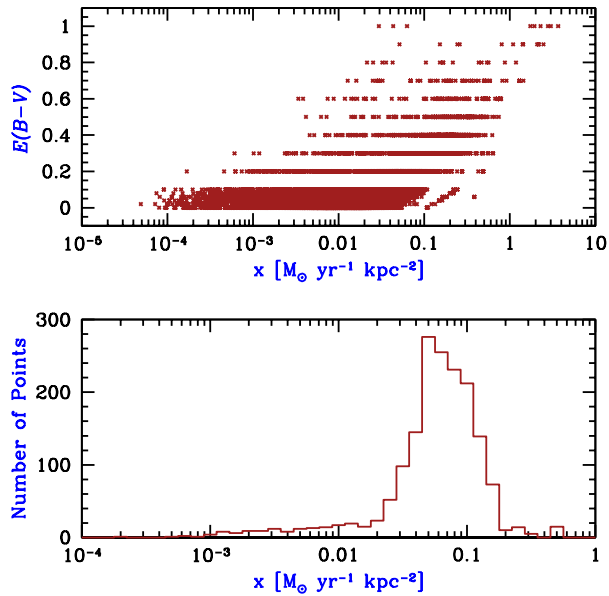}
\caption{Upper panel: Plot of the extinction $E(B-V)$ versus $x$ for the
redshift bin $z=0.7$--1.0. The detection threshold is apparent as a
linear cutoff at the upper-left end of the distribution of data points
(i.e., at low $x$ or high $E(B-V)$ values). Lower panel: Histogram of
the $x$ values in the upper panel that correspond to
$E(B-V)=0.2$. Note that the peak lies far from the detection limit.}
\label{fig:seleff}
\end{figure}

We now consider our main objective in this Letter which is to
correlate the extinction and the SFI. As noted in section~3, the
probabilities of the different combinations of extinction, redshift
and template for each galaxy can be calculated using the $\chi^2$
values associated with the combinations together with
equation~(\ref{eq:Pi}). Having obtained the probabilities, for each
galaxy we statistically sample this distribution 200 times in a Monte
Carlo approach. The data that are used in the following analysis are
the redshift, extinction and the calculated SFI values, for each of
the 200 sample combinations, for each of the 1927 galaxies used in
this Letter. This Monte-Carlo approach includes statistically all the
uncertainties and degeneracies from the photometric fits of the
individual galaxies. Figure~\ref{fig:seleff} (upper panel) plots the
extinction versus $x$ for combinations with redshifts
$z=0.7$--1.0. The binning of the extinction values is apparent. There
are 12,072 points in the plot and their distribution clearly points to
a correlation between the extinction and the SFI.

We note that some caution is needed here, since a correlation between
the extinction and the SFI would have been introduced into our
analysis simply as a selection effect; thus we must carefully
demonstrate that the observed correlation is instead a real effect. To
identify an object as a galaxy, the selection procedure (as described
in TWS) requires a minimum flux per pixel in the various filter
wavebands. At any given redshift, objects with very low x values or
with high extinction values may exist and not be detected. This
selection effect is evident in the upper panel of
Figure~\ref{fig:seleff} as a lack of galaxies at the low $x$ and high
$E$ region (i.e., the top-left corner). More specifically, the left
boundary of the collection of data points is quite linear (as is the
case in other redshift bins as well). Since the detectability of
galaxies is roughly set by a minimum surface brightness, and the
observed surface brightness is proportional to the intrinsic $x$ of
the galaxy times an extinction correction of the form $\exp[-a
E(B-V)]$ for some constant $a$, then galaxies with observed flux at
the detection limit should lie on a straight line in $E$ versus
$\log_{10}(x)$. 

In general, a correlation between the extinction and the SFI could be
inferred mistakenly due to this selection effect, but in
Figure~\ref{fig:seleff} this is not the case.  The selection effect is
relevant only at low $x$ and high $E$, but figure (3a) also shows a
clear lack of points at high $x$ and low $E$, where any galaxies would
be easily detected, and selection effects are absent. 

To demonstrate this more clearly, the data points at each extinction
value were tested to verify that most of the $x$ values lie far above
the detection limit, and thus the mean $x$ value (which we use below)
is insensitive to selection effects. The lower panel of
Figure~\ref{fig:seleff} shows a histogram of the $\log_{10}(x)$ values
that correspond to $E(B-V)=0.2$ in the upper panel of the same
figure. As can be seen, the detection limit in this case is around
$\log_{10}(x)=-3$, whereas most of the data points lie at
$\log_{10}(x)=-1.6$ or higher, and the average is at -1.1. This
behaviour, which we have verified for other extinction values and other
redshift bins, clearly indicates that the selection effect is not the
main cause of the correlation we find between the extinction and
SFI. We caution though that at the highest redshifts we consider,
there are fewer observed galaxies and fewer data points particularly
at high extinction, so the detection threshold affects the results
significantly in this regime. Additional data at fainter fluxes would
be very useful for checking the reliability of the high-redshift,
high-extinction points.

\begin{figure}
\includegraphics[width=84mm]{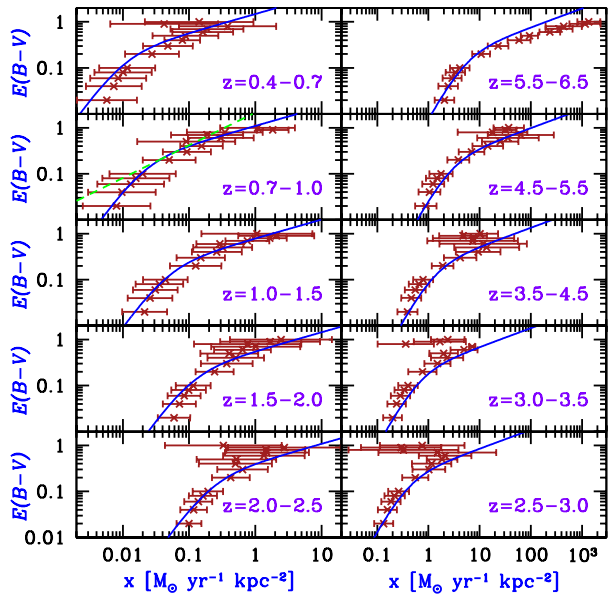}
\caption{Plots of the extinction $E(B-V)$ versus the SFI at various
redshift bins between $z=0.4$ and 6.5. The correlation between the
extinction and the SFI is clearly observed at all these epochs. We
show the double power-law fit to the data (solid curves). We also
compare to a power-law slope of 0.71 (dashed curve, $z=0.7$--1.0
panel). See section~5 for a discussion of the model fits}
\label{fig:many}
\end{figure}

In order to characterize the correlation between the extinction and
the SFI as a function of epoch, we divided the data into ten redshift
bins: 0.4--0.7, 0.7--1.0, 1.0--1.5, 1.5--2.0, 2.0--2.5, 2.5--3.0,
3.0--3.5, 3.5--4.5, 4.5--5.5 and 5.5--6.5. In each redshift bin, the
mean and standard deviation of all the $\log_{10}(x)$ values that
correspond to each specific extinction value were calculated. We
calculated symmetric errors here in order to use them below in a
$\chi^2$ minimization. The resulting extinction and mean
$\log_{10}(x)$ values, with 1-$\sigma$ error bars in $\log_{10}(x)$,
are plotted in Figure~\ref{fig:many} for all redshift bins. The
correlation between the extinction and SFI is clearly evident at all
epochs. Note that the errors in $x$ [which is proportional to
$(1+z)^4$] are smaller in the higher-redshift bins since the
photometric redshifts are more accurate due to the Lyman break (see
also figure~1).

\begin{figure}
\includegraphics[width=84mm]{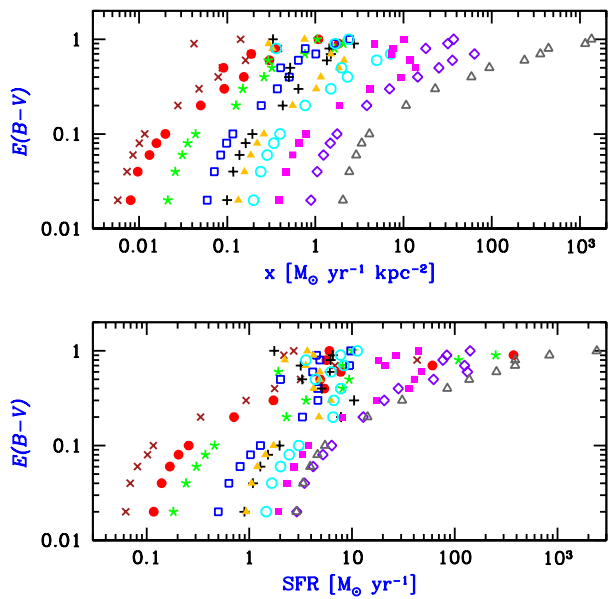}
\caption{Plot of $E(B-V)$ versus $x$ (upper panel) or versus the SFR 
(lower panel) for all redshift bins. The bins are $z=0.4$--0.7
($\times$), 0.7--1.0 ($\bullet$), 1.0--1.5 ($\star$), 1.5--2.0 (open
squares), 2.0--2.5 (+), 2.5--3.0 (solid triangles), 3.0--3.5
($\circ$), 3.5--4.5 (solid squares), 4.5--5.5 ($\diamond$) and
5.5--6.5 (open triangles).}
\label{fig:allz}
\end{figure}

Figure~\ref{fig:allz} (upper panel) plots all the redshift bins
together so that the evolution with redshift can be observed more
easily. It is evident that for a given SFI, the extinction decreases
with $z$. This behaviour is consistent with the increase in the
metallicity of the ISM with time \citep{i03}.  The lower panel of
Figure~\ref{fig:allz} is similar and is based on the same analysis as
the upper panel except that the extinction is plotted against the
total SFR and not against the SFI. Although the results in the lower
panel depend differently on the size distribution of galaxies, the
same behaviour is seen as in the upper panel. We further consider these
two different correlations in the following section.

\section{Fitting the Correlation}

In the previous section, the relation between the extinction and SFI
was described empirically. As noted in the introduction, this relation
is expected from the Schmidt law together with the correlation between
the gas and dust in the ISM of galaxies.

\citet{s59} put forth the hypothesis that the rate of star formation
in a given region varies as a power of the gas density within that
region. Observations yield direct estimates of surface densities and
thus, the relation between the SFI $x$ and the gas surface density has
been measured over a broad range of physical conditions in relatively
nearby galaxies, yielding the relation $x \propto \Sigma_{\rm
gas}^{1.4}$ \citep{k98}. The relation between the dust and gas content
in the ISM is less clear. If a constant dust-to-gas ratio is assumed,
then if the Kennicutt relation is valid at a given redshift it should
yield a linear relation between $x$ and $E$, with a slope of
$1/1.4=0.71$ in a log-log scale. However, this model is inconsistent
with the observational results shown in Figure~\ref{fig:many}. To
illustrate this, we show in the top-right panel ($z=0.7$--1) of this
figure a line with this slope (dashed curve). It is apparent in all
the redshift bins that the dependence of $E$ on $x$ in the actual data
is steeper than this power at low $x$ and shallower at high
$x$. Equivalently, the dust-to-gas ratio at both ends is lower than at
the central $x$ values, assuming the \citet{k98} relation is valid at
each redshift.

The behaviour in Figure~\ref{fig:many} suggests a double power-law
model; the precise form we choose is motivated by the conventional
model for the luminosity function of quasars [e.g., \citet{p97}]. In
addition, Figure~\ref{fig:allz} suggests a simple evolution with
redshift as a power of $(1+z)$, and so we fit a single model to all
epochs simultaneously:
\beq E(B-V)= \frac{E_0 (1+z)^{\gamma_1}} {\left[ \frac{x} {x_S 
(1+z)^{\gamma_2}} \right]^{-\alpha} + \left[ \frac{x} {x_S 
(1+z)^{\gamma_2}} \right]^{-\beta} }\ .
\eeq
This model consists of six free parameters: $E_0$ -- a normalization
parameter, $x_S$ -- the knee point that separates the two slopes,
$\gamma_1$ and $\gamma_2$ -- the powers of $(1+z)$ that describe the
redshift evolution of $E_0$ and $x_S$, respectively, and $\alpha$ and
$\beta$, which represent the asymptotic power-law slopes at low and
high $x$-values, respectively. Note that we excluded $E=0$ data points
from the fit and from Figures~\ref{fig:many} and \ref{fig:allz}.

We fit the model to all redshift bins by minimizing a combined
$\chi^2$ built according to the errors shown in
Figure~\ref{fig:many}. The figure shows that the model yields a rather
good fit to the data at all epochs. The redshift value of the fit
shown in each panel is the value at the center of the redshift
bin. The fitted parameters with their 1-$\sigma$ errors (which
correspond to a unit increase in the $\chi^2$ value from its minimum
value) are: $\log_{10}(E_0)=-0.65^{+.23}_{-.21}$,
$\log_{10}(x_S)=-2.79^{+.19} _{-.23}$, $\gamma_1=0^{+.03}_{-.04}$,
$\gamma_2 = 4.14^{+.12} _{-.13}$, $\alpha=2.03 ^{+.71} _{-.42}$, and
$\beta=0.41 ^{+.11}_{-.13}$. 

The best-fit value $\gamma_1=0$ implies a relation of constant shape
that simply shifts horizontally with redshift. The $\chi^2$ value of
the best fit is 111.6 while the number of degrees of freedom is 134,
which suggests that the errors may be somewhat overestimated, which is
perhaps related to our assuming uncorrelated errors for the various
points when constructing the $\chi^2$ for this fit.
Figure~\ref{fig:allz} reveals another possible characterization of the
correlation by relating the extinction to the SFR instead to the
SFI. The double power law was fitted to these data as well, using the
same analysis as described above for the case of the SFI. The fitted
parameter results in this case, with a best-fit $\chi^2$ value of
107.1, are: $\log_{10}(E_0)=-0.39^{+.08}_{-.07}$,
$\log_{10}(x_S)=-0.62^{+.14} _{-.13}$, $\gamma_1=0^{+.08}_{-.53}$,
$\gamma_2 = 2.16^{+.21} _{-.20}$, $\alpha=2.17 ^{+.42} _{-.3}$, and
$\beta=0.22 ^{+.05}_{-.06}$.

The two fits presented here depend on the sizes of galaxies in
different ways. While the SFI is a more local quantity, the total SFR
depends also on the global size of the galaxy. A galaxy's overall size
and mass affects star formation through global feedback effects. In
particular, supernova feedback is efficient only in relatively small
galaxies which results in various observed correlations in the
properties of galaxies at low redshift [e.g., \citet{db05}]. 
Additional data for a larger number of galaxies is required to fully
explore the effect of the size of a galaxy on its average extinction
value.
    
\section{Conclusions}

We have shown that a clear correlation exists between the extinction
of galaxies and their SFI. This correlation is apparent at various
epochs ranging from z=0.4 to z=6.5 and is in general agreement with an
increasing metallicity in the ISM with time. We have also shown that a
similar correlation remains when the galaxies' SFR is considered
instead of their SFI, a case in which global effects such as
supernovae feedback may affect the behaviour. We have fitted a double
power-law model to the results and shown that it fits the data
well. We have suggested that this correlation is a natural consequence
of the Schmidt law, which relates the SFI of galaxies with their gas
content in the ISM, together with a relation that exists between the
gas and dust content in the ISM. The exact shape of the correlation
should be further investigated both to further eliminate selection
effects and to shed more light on the cosmic evolution of gas and dust
in galaxies.

\section*{Acknowledgments}
RB acknowledges support by Israel Science Foundation grant 28/02/01.

\bsp

\label{lastpage}

\end{document}